# Ion-beam-assisted characterization of quinoline insoluble particles in nuclear graphite

Qing Huang[1,*], Xinqing Han[2], Peng Liu[2], Jianjian Li[1], Guanhong Lei[1], Cheng Li[1]

[1]Shanghai Institute of Applied Physics, Chinese Academy of Sciences, Shanghai 201800, China

[2]School of Physics, State Key Laboratory of Crystal Materials, Key Laboratory of Particle Physics and Particle Irradiation (MOE), Shandong University, Jinan 250100, China

[*]Corresponding author. *E-mail address: huangqing2012@sinap.ac.cn*

**Abstract:** The irradiation behavior of graphite is essential for its applications in nuclear industry. However, the differences between graphite's behaviors are not well understood because of the very limited knowledge of microstructural differences between graphites. One typical structure, quinoline insoluble (QI) particle, was investigated using IG-110 and NBG-18 graphite. After irradiation, the QI particles on the polished surface were proved to become hillocks and can be easily identified by using a scanning electron microscope (SEM). Thus a method combining ion irradiation and SEM characterization was proposed to study the distribution and concentration of QI particles in graphite. During irradiation, the QI particles were found to evolve into densified spheres which have weak bonding with the surrounding graphite structures, indicating that the densification of QI particles does not contribute obviously to graphite dimensional shrinkage. A much higher concentration of QI particles in NBG-18 than IG-110 was characterized and is suggested to be responsible for the smaller maximum dimensional shrinkage of NBG-18 than IG-110 during irradiation.

# 1. Introduction

After being used in fission reactors for over 70 years, graphite is still proposed to be the neutron moderator and neutron reflector material in two types of Generation IV reactors: molten salt reactors and very high temperature reactors. The goals of generation IV reactors are to improve the fission reactors in terms of sustainability, safety and reliability, economics, proliferation resistance and physical protection. Structural materials for Generation IV reactors usually serve in a high temperature and intense neutron irradiation environment. The irradiation behavior of one graphite material has to be evaluated to see whether it is qualified for neutron moderator or reflector in Generation IV reactors.

Nuclear grade graphite is known to be high-purity, polycrystalline and near-isotropic. Graphite is usually produced from a coke (petroleum coke or pitch coke) and pitch binder. Even though the graphitization temperature for a nuclear graphite product could be as high as 3000 ºC, typical structures in the raw materials leave their marks in the final product (such as the filler sizes and the concentration of quinine insoluble particles). Given the different sources of raw materials and various processing technologies that adopted by manufacturers, commercial nuclear graphite grades show very different properties, also different behaviors during neutron irradiation.

Neutron irradiation produces displaced atoms in nuclear graphite, leading to an expansion in the $c$ axis and a shrinkage in graphite basal planes. Countless graphite crystallites in artificial graphite material evolve and interact with each other and adjacent porosity, resulting in the changes of graphite volume, mechanical and thermal properties. Based on the irradiation data that obtained from neutron irradiation experiments in the past, the dimensions of nuclear graphite usually have a tendency of decreasing first and then increasing [1,2]. The maximum dimensional shrinkage and the doses at which the dimensional change reaches the maximum and returns to zero are different for different nuclear graphites.

The historical nuclear grades of graphite are not available now. Therefore, given the detailed irradiation data of historical graphite grades, nowadays irradiation programs of commercial grades of graphite for their applications in the very high temperature reactors are still going on [1,3]. These irradiation programs include many grades of graphite from different countries and intend to reach high doses at various temperatures. Although they are costly and time-consuming, these programs are necessary today and extremely important for the development of next generation graphite moderated reactors. In the case of molten salt reactors, the molten salt is in direct contact with graphite components. In order to prevent salt penetration, the graphite pores should have sizes smaller than 1 μm, which may be achieved by using a ultra-fine grained coke filler [4,5]. The development of a suitable graphite for molten salt reactors is still ongoing. The irradiation behavior of graphite that developed for molten salt reactors has not been revealed.

It was said that the size and cost of a graphite irradiation program "*can be significantly reduced through understanding the relationships between the graphite microstructure and properties*" [6]. Furthermore, given the need of simulating and predicting nuclear graphite's behavior, typical microstructures and their responses to irradiation should be fully understood. There have been many studies on graphite microstructures that irradiated with neutrons [7-10], electrons [11-13] and ions [14-16]. Irradiation-induced effects on the atomic level, including decreasing of the $sp^2$ content, fragmentation and distortion of graphite basal planes, decreasing of *a*-lattice parameter, increase of *c*-lattice parameter, dislocation dipole and climb of dislocations in graphite lattice, were reported. It was suggested that irradiation induced effects in graphite may be considered to have two components: "bulk microstructural effects" and "intrinsic lattice effects" [7]. The "intrinsic lattice effects" shown above are apparently common effects in all graphite materials. It is the so called "bulk microstructural effects" that lead to the divergent behaviors of many nuclear graphites during irradiation. However, the microstructural difference between graphites has not been well studied. Karthik et al., studied the microstructures of three graphites (NBG-18, IG-110 and PCEA) using a transmission electron microscope (TEM) and expressed that NBG-18 contains a higher concentration of quinine insoluble (QI) and chaotic structures than the other two graphites [17]. The QI particle, once referred to as rosettes, is one typical structure in nuclear graphite. The QI particles have a roughly spherical appearance and contains many packets of graphite sheets that were arranged concentrically [18]. The TEM has usually been adopted to characterize the QI particles in nuclear graphite [7,17]. Although the concentration of QI particles was reported to vary from graphite to graphite, the QI particles' influence on irradiation behaviors of nuclear graphite is not revealed. In this study, two nuclear graphite grades were irradiated with energetic xenon ions at high temperature. The evolution of QI particles during irradiation were observed and explained, which provides a new way to determine the distribution and concentration of QI particles in graphite. After irradiation to a high dose, the structures including QI particles were characterized. The QI particles' influence on graphite's volume changes during irradiation was discussed.

## 2. Materials and Methods

### 2.1 Materials

Two graphites were used in this study. The IG-110 graphite, from Toyo Tanso Inc., is produced from a petroleum coke with an average grain size of ~20 μm. The NBG-18 graphite, from SGL Group, is manufactured from a pitch coke with an maximum grain size of 1600 μm. According to the standard ASTM D7219, NBG-18 and IG-110 are regarded as medium-grained graphite and superfine-grained graphite respectively. Typical properties of the two graphite grades are listed in Table 1.

Table 1 Typical properties of IG-110 and NBG-18 graphite

|        | Density ($g/cm^3$) | Compressive strength (MPa) | Tensile strength (MPa) | Flexural strength (MPa) | Coefficient of thermal expansion ($10^{-6}/K$) | Thermal conductivity (W/m/K) | Young's Modulus (GPa) |
|--------|--------|--------|--------|--------|--------|--------|--------|
| IG-110 | 1.77 | 78 | 25 | 39 | 4.5 | 120 | 9.8 |
| NBG-18 | 1.85 | 80 | 20 | 30 | 4.5 | 140 | 11.5 |

Firstly, samples of the two graphite grades were ground using 1500 grit emery paper in order to remove the entire original rough surface. Then the samples were polished successively on cloth using 3000 nm and 50 nm diamond suspensions. It is worth mentioning that mechanical polishing produces surface defects, resulting in an increase of D peak intensity in the Raman spectrum. However, a flatten surface could be formed by polishing and is suitable for analyzing the evolutions of filler particles, pores and other microstructures during irradiation. The polished samples were cut into small slices (length × width × thickness: $7 \times 1.5 \times 1$ $mm^3$) for irradiation.

### 2.2. Ion irradiation

Two samples, one is IG-110 and the other is NBG-18, were irradiated by (7 + 4 + 2) million electron volts (MeV) xenon ions. Irradiation was performed at the 320 kV High voltage Experimental Platform in the Institute of Modern Physics, Chinese Academy of Science. Twenty six electrons were stripped off the xenon atoms ($Xe^{26+}$) in order to reach the energy of 7 MeV, while twenty electrons were stripped off ($Xe^{20+}$) in order to reach the energy of 2 and 4 MeV. Xenon ions were adopted in this study because they are chemically inert and can create displaced carbon atoms in a high speed. Multi-energy xenon ions were chosen to irradiate the samples in this study in order to produce a uniformly damaged layer in samples. The xenon ion beam was scanned in two directions. An annular Faraday cup was used as an aperture in front of the target, so that the target could be irradiated uniformly by the xenon ions. The sample was fixed on a metallic plate which faces the xenon ion beam. The metallic plate was installed in the middle of a circular heater, and the temperature of the sample was maintained to 600 °C during irradiation. The temperature of the sample was measured and monitored by a thermocouple. There is another Faraday cup behind the target, in order to measure the ion beam passing through the aperture. When the target has been mounted and is ready for irradiation, the ion beam is blocked by the target and cannot be measured by this Faraday cup. Therefore, the current of the ion beam was monitored during irradiation by the annular Faraday cup in front of the target. The beam currents of the xenon ions were recorded to be in a range from 1.6 to 2.6 µA. A high vacuum (pressure < $5 \times 10^{-4}$ Pa) in the irradiation chamber was kept during heating and irradiation.

The xenon ions irradiation process was simulated by using "The Stopping and Range of Ions in Matter" (SRIM) program. A value of displacement threshold energy ($E_d$) has to be given for a proper SRIM simulation. When evaluating the neutron irradiation induced displacements in carbon materials, the value of $E_d$ was reported to be 31 eV [19]. A theoretical study reported that the $E_d$ has a value of 25 eV at the room temperature [20]. At a temperature of 900 K, the value of $E_d$ increases to 30 eV [20], which is very close to the value used for neutron irradiation damage calculation. Therefore, a value of 31 eV was applied for $E_d$ in this study. It is worth mentioning that changing the value of $E_d$ only alters the calculated number of displacement atoms and it does not influence the profile of the displacement atoms.

The SRIM simulations with two modes were tried to calculate the number of displacement atoms. One mode is "Ion Distribution and Quick Calculation of Damage" and the other mode is "Detailed Calculation with full Damage Cascades". The number of displacements in nickel and iron calculated by using these two modes were compared with that derived from the well-known model developed by Norgett, Robinson and Torrens (NRT model) [21]. It was found that the SRIM simulation with the "Quick Calculation mode" produces displaced atoms in nickel and iron comparable to the NRT model while the "full Damage Cascade mode" produces much more displaced atoms [22]. Therefore the SRIM calculation with the "Quick Calculation mode" has been recommended. However, in the case of xenon ion irradiation into graphite, the numbers of displaced atoms calculated by using SRIM program with the two modes are comparable and are 1/3 less than that derived from the NRT model. In this study, inert xenon ion irradiation at high temperature was applied to graphite in order to simulate the neutron irradiation damage which has been calculated by using the NRT model [19]. Therefore, the number of displaced atoms that derived from the NRT model were used and are 19180, 14916 and 10026 displacements/ion for 7, 4 and 2 MeV xenon ions, respectively. Then, the profile of displacements per atom (dpa) was derived from the ion fluence (ions/cm$^2$). The dose was accumulated to 4, 15 and 25 dpa with 1 dpa equals to $(8.6+3.44+2.58)\times 10^{14}$ ions/cm$^2$.

### 2.3. Characterization techniques

Thin foils of un-irradiated graphites were prepared for TEM observation. Samples with diameters of ~3 mm were cut from the un-irradiated graphite materials and then were ground to disks with thicknesses of ~50 μm. A dimple grinder was used to further grind the center of the disks. Then, a penetration hole was formed in the disks by low-energy (3 - 5 keV) argon ion milling using a Gatan Model 691 system. A 200 kV FEI Tecnai G2 F20 TEM was used to characterize the QI particles in the two graphite grades. In addition, the NBG-18 sample was heated to 600 °C during TEM observation for in situ electron irradiation. The electron beam (200 kV) with a diameter of ~100 nm was focused on one QI particle. The beam current is ~15 nA. The image of the QI particle was captured every 2 minutes during irradiation. The electron fluences were accumulated to $1.4 \times 10^{23}$, $2.8 \times 10^{23}$, $4.2 \times 10^{23}$ and $5.6 \times 10^{23}$ electrons/cm$^2$.

One specimen for TEM characterization was also prepared for the NBG-18 graphite sample irradiated to 25 dpa, in order to investigate the structures of irradiated QI particles. To observe the cross section of the irradiated sample, two slices were cut from the xenon-ion-irradiated NBG-18 sample. The irradiated surfaces of the two slices were glued face to face. The edges of the specimen were milled until it can be inserted into a 3-mm-diameter copper tube. Then the tube was cut into small pieces. The remaining processing steps with the small pieces are the same as those mentioned above.

Surface of the polished samples was observed using a scanning electron microscope (SEM) before ion irradiation and after irradiation to 15 and 25 dpa, in order to study the surface morphology changes and QI particle evolution during irradiation.

The Raman scattering spectra were collected from the NBG-18 graphite before irradiation and after irradiation at each dose. A Bruker SENTERRA confocal Raman microscope was used to measure the Raman spectra at room temperature. The excitation beam has a wavelength of 532 nm.

A simplest fit with two symmetric Lorentzians is often used for the Raman spectra of crystalline graphite structures. But two peak fitting is not suitable for the Raman spectra of highly disordered carbon materials. Typically, a fit with four symmetric peaks (D peak, G peak and other two peaks) involving the Gaussian lines has been used for the Raman spectra of highly disordered carbon materials. However, four Gaussian lines usually can provide good fitting with more than one possibilities for the Raman spectra of highly disordered graphite. A unique and appropriate four peak fit cannot be distinguished from all the possibilities, which may cause inconsistency in interpreting the Raman spectra of irradiated graphite within literatures.

Instead of using a four peak fitting, the Raman spectra were fitted with a symmetric Lorentzian for the D peak and an asymmetric Breit-Wigner-Fano (BWF) line for the G peak in this study [23]. Because of the peak broadening induced by irradiation, the small D' peak emerged in the Raman spectrum of as-polished graphite was not recognizable after xenon ion irradiation. Therefore the fitting of the D' peak was not conducted.

## 3. Results

### 3.1 The QI particles in graphite before irradiation

Fig. 1 shows a typical QI particlein IG-110 graphite. There are many small packets of graphite sheets in QI particles. Empty voids were found in between the packets. The *c* axis of these packets is directed radially towards the center of the QI particles [18]. Usually, the QI particles are surrounded by folded graphite structures which are shown outside the dashed circle in Fig. 1.

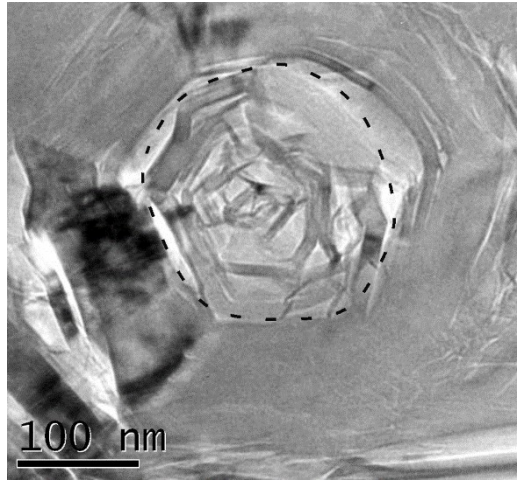

Fig.1 TEM micrograph of one typical QI particle (indicated by a dashed circle).

TEM micrographs of QI particles in NBG-18 and IG-110 graphite are shown in Figs. 2(a) and 2(b) respectively. It can be seen that the QI particles tend to agglomerate in nuclear graphite. The concentration of QI particles in IG-110 is much less than that in NBG-18 graphite, which is consistent with the conclusion of the previous study [17].

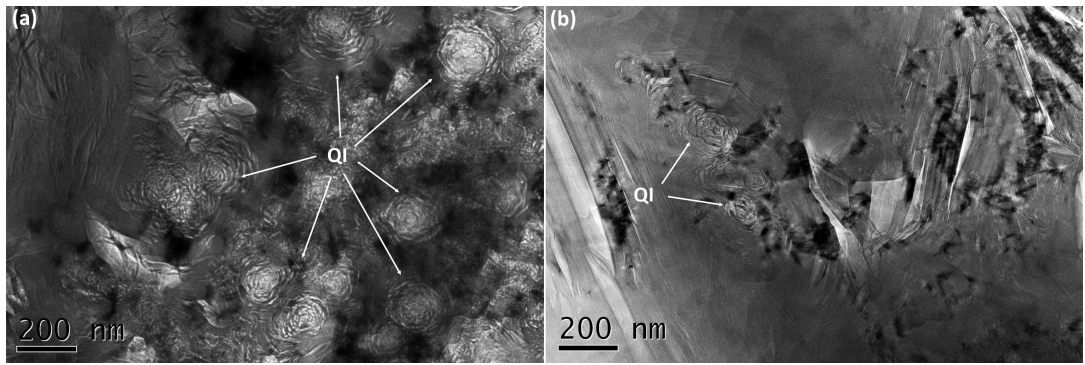

Fig. 2 TEM micrographs of microstructures with QI particles in (a) NBG-18 and (b) IG-110.

### 3.2 Identifying the QI particles on irradiated graphite surface

The SEM micrographs of irradiated surfaces of NBG-18 and IG-110 are shown in Figs. 3(a) and 3(b) respectively. The polished surface became rough after irradiation. Wrinkle-like textures (marked by W) are found on both graphites' surface. These wrinkle-like textures were firstly identified on needle-like fillers in IG-110 graphite after argon ion irradiation [14]. Therefore, they are believed to be the graphite structures which have the long range order. The "wrinkles" show the directions of curved graphite sheets. The size of the wrinkle-like textures represent the size of these graphite structures.

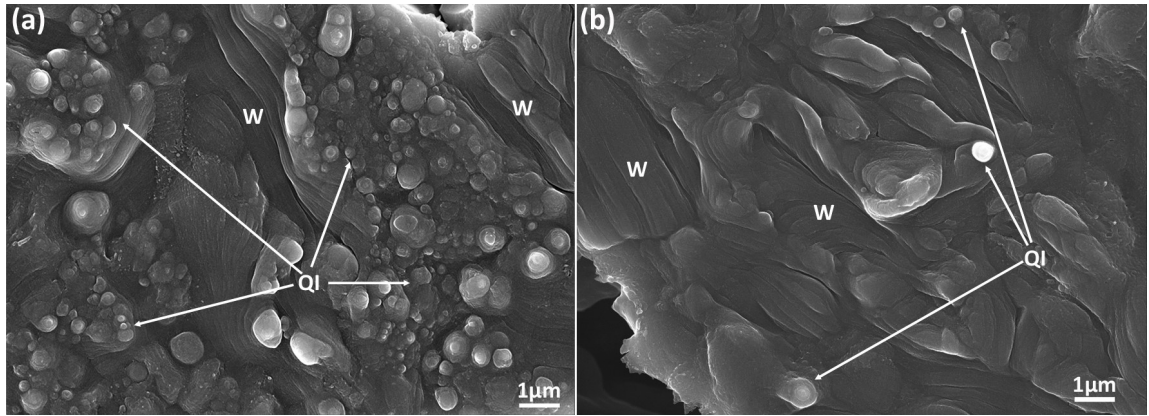

Fig. 3 SEM micrographs of polished surfaces of (a) NBG-18 and (b) IG-110 after irradiation to 15 dpa. The "wrinkle-like" structures shown on irradiated surface are marked by "W".

The main difference between the two graphites' surface morphology is the concentration of small round hillocks (indicated by arrows in Fig. 3). These hillocks have very small sizes and tend to agglomerate, especially in NBG-18. Therefore, they are considered to be the QI particles. In order to prove that, two QI particles were identified on the as-polished surface of the NBG-18 sample (Fig. 4(a)). Typical feature of QI particles, a loose cores surrounded by folded graphite sheets (indicated by two dashed circles), can be clearly seen in Fig. 4(a). After irradiation, two hillocks emerged at the same place. It can been seen that the hillocks, especially the left one, shrink radially and grow in height during irradiation. Faint round wrinkles are found on the two hillocks.

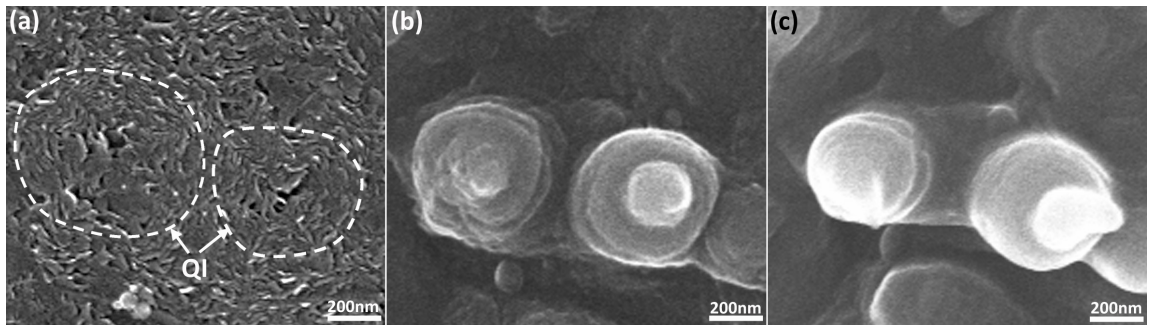

Fig. 4 SEM micrographs showing two QI particles evolve into hillocks. (a) Two QI particles indicated by dashed circles on the as-polished surface, (b) after irradiation at 15 dpa, and (c) after irradiation at 25 dpa.

In order to show the distribution of QI particles in NBG-18 graphite, Fig. 5 shows a SEM micrograph of the irradiated NBG-18 surface. The entire irradiated surface is occupied by agglomerations of hillocks, along with the wrinkle-like structures. Large cracks shown in Fig. 5 is typical for NBG-18 graphite's filler and were formed during calcination of the filler coke. This kind of calcination cracks are also present in fillers of Gilsocarbon and PCEA, other two medium-grained graphites [18, 24]. Therefore, the Fig. 5 shows an area in one typical filler of NBG-18 graphite. It is interesting to see a large number of QI particles in filler since the QI particles were always reported to be formed in the binder phase [7,18]. Based on the morphology of irradiated NBG-18 surface, it can be deduced that the filler in NBG-18 graphite is composed of many graphite structures (represented by the wrinkle-like structures with sizes from several microns to tens of microns) which are separated by agglomerations of QI particles (represented by the hillocks).

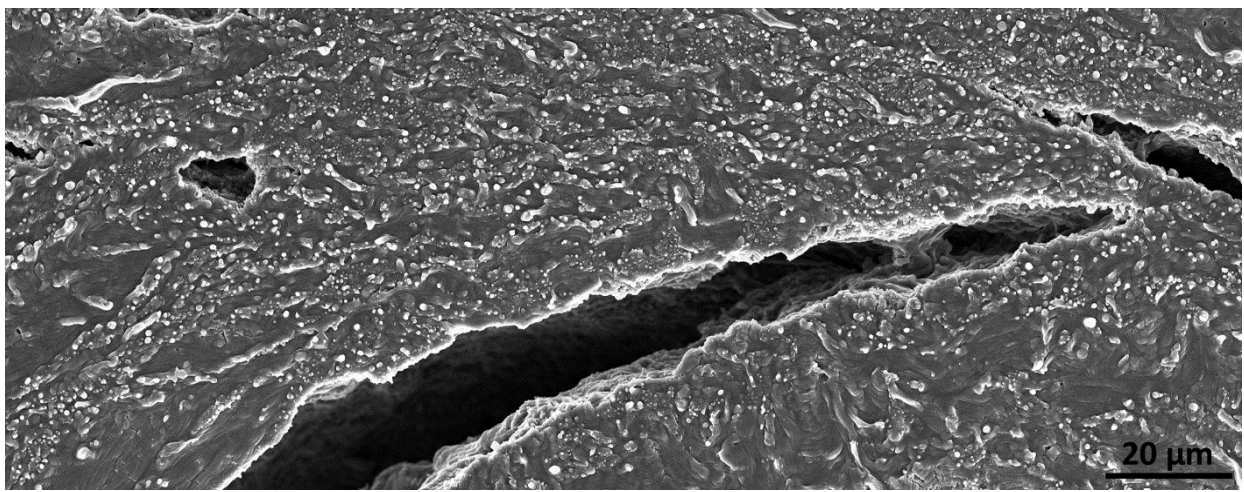

Fig. 5 SEM micrograph of irradiated NBG-18 surface showing calcination cracks and a large number of hillocks.

### 3.3 The Raman spectra of irradiated graphite

The Raman spectra were measured at the regions of agglomeration of QI particles on polished NBG-18 surface. Fig. 6(a) shows the Raman spectra before and after xenon ion irradiation. The irradiation doses are also given in Fig. 6(a). Three peaks (The G mode appearing at 1582 cm$^{-1}$, The D mode located at ~1350 cm$^{-1}$ and the small peak D' at 1620 cm$^{-1}$) emerged in the spectrum of un-irradiated graphite. During xenon ion irradiation, the graphite structures were damaged gradually by nuclear collisions, resulting in peak broadening and overlapping.

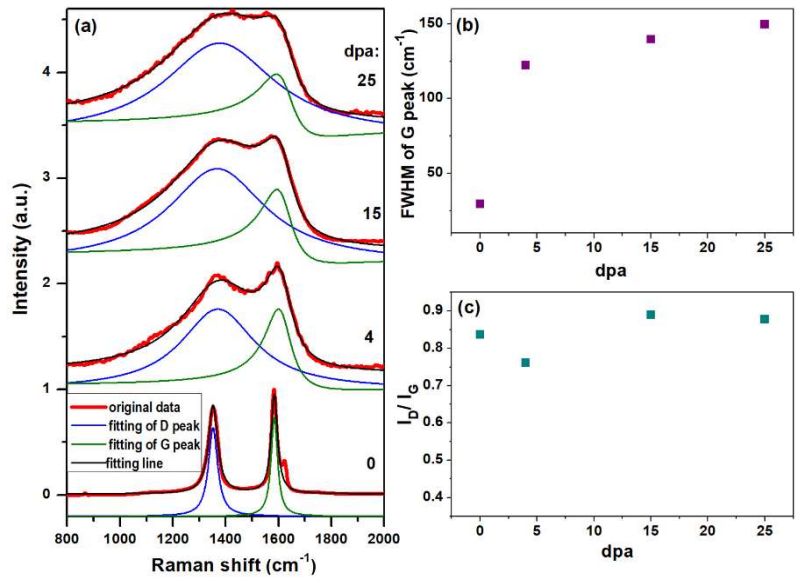

Fig. 6 (a) The Raman spectra and fitting curves of agglomeration of QI particles before and after xenon ion irradiation to three doses. (b) The full width at half maximum (FWHM) of the G peak at various doses. (c) $I_D/I_G$ at various doses.

Figs. 6(b) and 6(c) show the full width at half maximum (FWHM) of the G peak and the ratio of the intensities of D and G peaks ($I_D/I_G$) respectively. The FWHM of G mode of the as-polished NBG-18 sample is ~30 cm$^{-1}$. During xenon ion irradiation at 600 °C, the FWHM increased monotonically to a value of ~150 cm$^{-1}$ (25 dpa). It was reported that the irradiation induced disordering of the graphite lattice might cause a strong coupling between the discrete G mode and a continuum of phonon states, resulting in the broadening of the G peak [25].

The D mode was explained as a "breathing mode" of sp$^2$ carbon atom rings and is forbidden in a perfect graphite lattice [9]. In-plane defects have to be introduced into graphite to excite the D peak. The region of agglomeration of QI particles contains many small graphitic structures with abundant defects in boundaries. Therefore the $I_D/I_G$ ratio was high (0.84) even without irradiation. At the beginning of irradiation, the number of in-plane defects increased but enough carbon atom rings preserved, resulting in an increase of D peak intensity with limited overlapping of peaks [26]. It has been reported that the $I_D/I_G$ ratio increase fast and reaches a maximum of ~1.4 at quite low doses (<0.02 dpa) for graphite irradiated at room temperature [9,15,26]. Along with the increase of $I_D/I_G$ ratio, the graphite structure evolved into nano-crystalline graphite. With further irradiation, the lattice disordering increased and the number of ordered rings decreased, resulting in a decrease of D peak intensity and broadening of Raman scattering peaks. Fig. 6(a) shows that the Raman peaks at 4 dpa are already highly broadened, so that the maximum of $I_D/I_G$ ratio should appear at lower doses than 4 dpa. It is worth mentioning that an inversely proportional relationship between the $I_D/I_G$ ratio and the in-plane crystallite size was proposed and has been used to characterize the crystallinity of graphitic structures [27]. However, this relationship can only be used to calculate the crystallite size in the increasing stage of $I_D/I_G$ ratio[23]. At higher irradiation doses, the $I_D/I_G$ ratio decreases and this relationship is apparently not valid any more.

Both the G peak width and the $I_D/I_G$ ratio tended to be stable at doses higher than 15 dpa, indicating that the graphite structure at high irradiation doses, although damaged, is very stable. It was reported that the $I_D/I_G$ ratio decreased toward 0.2 in an amorphous carbon material [23,28]. Fig. 6(c) shows that the values of $I_D/I_G$ ratio at high doses are around 0.9, indicating that graphite is not amorphous after xenon ion irradiation to 25 dpa at 600 °C.

### 3.4 Irradiation effects in QIs in nuclear graphite

One QI particle was in-situ irradiated by the electron beam in TEM at 600 °C. Figs. 7(a) - 7(c) show the QI particle before irradiation and after irradiation for 4 and 8 minutes (with electron fluences of $2.8 \times 10^{23}$ and $5.6 \times 10^{23}$ electrons/cm$^2$) respectively. During irradiation, the small packets of graphite sheets swelled in the $c$ axis and shrank in the basal plane. The $c$-axis swelling was readily absorbed by the voids, and the shrinkage in basal plane resulted in the densification of the QI particle. The black arrows show the evidence of shrinkage of the QI particle. Densification of QI particles has also been reported in neutron irradiated NBG-18 [7].

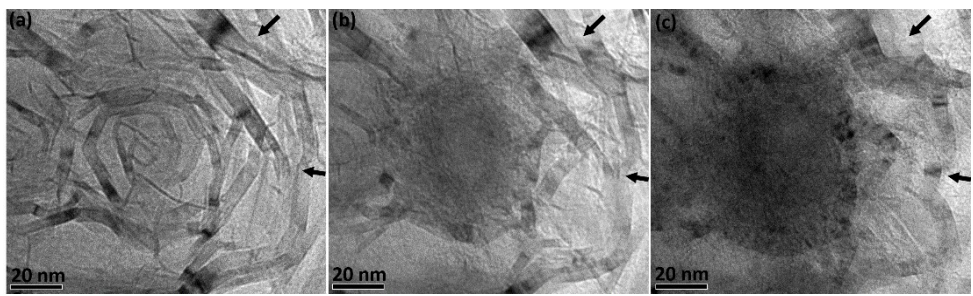

Fig. 7 Evolution of one QI particle in NBG-18 during in-situ electron beam irradiation at 600 °C. Electron dose: (a) 0, (b) $2.8 \times 10^{23}$ and (c) $5.6 \times 10^{23}$ electrons/cm$^2$.

Fig. 8(a) shows the TEM micrograph of NBG-18 irradiated to 25 dpa. This TEM sample was processed not only by the xenon ion irradiation but also by subsequent low energy argon ion milling during TEM sample preparation. It should be careful to distinguish the damage induced by xenon ion irradiation from that induced by low energy argon ion milling. Fig. 8(a) shows a highly damaged surface layer with a thickness of ~3.1 μm which is consistent with the xenon ion range (3 μm) obtained from the SRIM simulation. Therefore, the damaged surface layer is induced by xenon ion irradiation. The boundary (indicated by a white dashed line) between the xenon-ion-irradiated layer and the un-irradiated substrate was clearly seen in the TEM micrograph. The structures in the substrate were not irradiated by xenon ions but subjected to low energy argon ion milling. Typical QI particles and graphite structures were found in the un-irradiated substrate, indicating that the damage induced by low energy argon ion milling could be ignored compared to the damage induced by high dose xenon ion irradiation.

Many densified QI particles with a spherical appearance are shown in the irradiated layer. Two areas indicated by the dashed squares are magnified in Fig. 8(b) and 8(c). Irradiated and un-irradiated QI particles are identified in Fig. 8(b). The loose core of QI particles and the surrounding folded graphite sheets shrank and evolved into the densified spheres. The densified spheres were found to have week bonding with the adjacent structures.

There are plenty of lenticular cracks with widths of several nanometers or tens of nanometers shown in Figs. 8(a) - 8(c). Lenticular cracks are typical structures in nuclear graphite and are known as Mrozowski cracks. During cooling down from the temperature of graphitization, cracks opened between graphite basal planes by anisotropic thermal shrinkage. Both heating and irradiation could cause a remarkable expansion of graphite structure in the $c$ axis, which can be readily absorbed by these cracks. Thank to these lenticular cracks, graphite has a good resistance to thermal shock and irradiation induced swelling and disintegration. After irradiation to 25 dpa, graphite structures were expected to swell in the $c$ axis and close the lenticular cracks. But Fig. 8 shows that the graphite structures between the QI particles still show plenty of cracks lying between the graphite sheets because of the shrinkage of the adjacent QI particles.

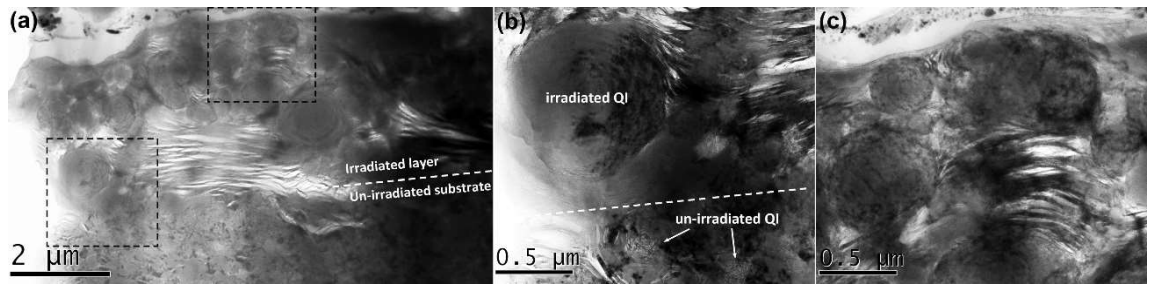

Fig. 8 (a) TEM micrograph of the NBG-18 sample irradiated to 25 dpa. The white dashed line indicate the end of ion range. The areas indicated by two black dashed squares are magnified in (b) and (c).

## 4. Discussion

Since the "hillock" structures were formed on the irradiated surface, they are easily to be regarded as results of sputtering under xenon ion bombardments. The SRIM simulations using the "Monolayer Collision Steps/Surface Sputtering" mode were conducted. The sputtering yield of xenon ions (normal incidence) with energies of 2, 4 and 7 MeV were obtained from SRIM simulation and are 1.63, 1.35 and 0.92 atoms/ion respectively. In total, irradiation to 25 dpa sputtered $4.19 \times 10^{16}$ atoms/cm$^2$ corresponding to only a 4.6-nm thick graphite layer being stripped off. Therefore, the "hillock" structures emerged on the irradiated graphite surface cannot be produced by sputtering under xenon ion bombardments.

An explanation of the emerging of hillocks is shown in Fig. 9. Fig. 9(a) shows a sketch of a QI particle which happened to be at the polished surface. Therefore, the surrounding graphite sheets are cut at the polished surface. During irradiation, along with the shrinkage of the QI particle, the fractured graphite sheets shrink in basal plane and swell along the $c$ axis. At the beginning of irradiation, the $c$-axis swelling is accommodated by microcracks. The folded graphite sheets surrounding QI particles (Fig. 1) have very narrow (< ~20 nm) and short (< ~100 nm) microcracks which can be readily closed by fast $c$-axis swelling. After the microcracks closed, the structure grow along the $c$ axis. Meanwhile, the folded graphite sheets shown in Fig. 9(a) shrink in the basal plane, resulting in a hillock shrinking radially and growing in height during irradiation. Experimental and theoretical studies [13,29] show that irradiation induced fragmentation of graphite sheets, which is also indicated by the segmented lines shown in Fig. 9(b).

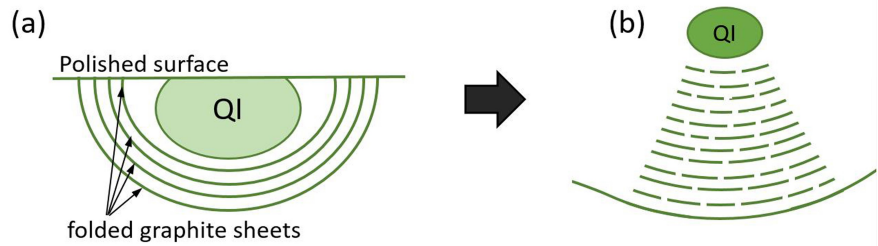

Fig. 9 Mechanism of the hillock's formation on the irradiated surface. (a) a QI particle at polished surface before irradiation, (b) the QI particle after irradiation.

TEM has been used to characterize the QI particles in nuclear graphite. This is believed to be the first report of SEM characterization of single QI particle and distributions of QI particles in nuclear graphite. For TEM observation, the field of view is limited to several μm$^2$ when observing the QI particles in graphite. On the contrary, the hillocks can be easily identified in a scanning electron microscope at magnifications around 400× with the field of view much larger than 10000 μm$^2$, which shows the global distribution of QI particles in one graphite. When comparing the concentration of the QI particles between different graphites, the SEM characterization of the hillocks is believed to be more convincing than the very local TEM characterization. The SEM micrographs of irradiated graphite clearly show that NBG-18 contains a large number of QI particles with a concentration much higher than that in IG-110.

At last, the QI particles' influence in graphite's volume changes during neutron irradiation is discussed. However, instead of neutron irradiation, ion irradiation was adopted to study the QI particles' evolution (Fig. 8). Ion beam irradiation and electron beam irradiation has long been applied instead of the time-consuming and expansive neutron irradiation in studying the irradiation effect in microstructures of structural materials. In the case of high-purity nuclear graphite, the changes of mechanical and thermal properties induced by neutron irradiation are outcomes of atom displacement damage which can also be produced much faster by ion irradiation. Energetic neutrons and ions have different track in graphite. Neutrons barely interact with electrons and lose its energy by nuclear collision, leading to a very deep track in the target. Ions interact with both electrons and nuclei and only leave a very short track in the target. The neutrons could create primary knocked-on atom (PKA) in the entire sample (centimeter size), while the ions only create PKA at the sample's surface (micron size). However, after being produced by ions or neutrons, the PKAs are ions and create the most part of displaced atoms in the target. Therefore, similar effects are expected for ion irradiation and neutron irradiation. Actually, the well-known neutron irradiation effects, such as irradiation induced dislocations, d-spacing increase, lenticular crack closing, shrinkage in basal plane and swelling in the *c* axis, have been reported in ion irradiated graphite [14,30]. Creep of graphite induced by proton irradiation was reported to simulate the in-reactor creep of graphite [31]. Nano-indentation tests were performed on nuclear graphite samples irradiated by carbon ions or protons and shows an increase of modulus which is consistent with neutron irradiation effects in graphite [26,32]. Therefore, the ion beams are suitable for studying irradiation effects in graphite microstructures. It must be noted that swift heavy ions should be avoided when simulating neutron irradiation effects in nuclear graphite. Very high electron excitation could be produced by swift heavy ions and leaves a disordered ion track in the graphite lattice [33]. The ion's energy that transferred to target's electrons has to surpass a threshold value in order to form an observable ion track. A threshold value of 730 eV/angstrom has been reported for graphite [33]. By SRIM simulation, the peak value of electron excitation caused by 7 MeV xenon ions is 250 eV/angstrom which is smaller than the threshold value. Disordered ion tracks cannot be formed by electronic energy deposition in this study. Xenon ions used in this study are believed to create defects in graphite only by nuclear collisions, in the same way that happened during neutron irradiation.

All the microstructures (such as long-range ordered graphite structure, microcracks, QI particles and others) could affect the bulk property changes during irradiation. Although the NBG-18 was manufactured with fillers much larger than those used for IG-110, this study shows that the long-range ordered graphite structures (indicated by "wrinkles" shown on irradiated surface) within the two graphite grades have similar sizes. On the sub-micron scale, we did not found obvious difference in microcracks between the two graphite grades by using a TEM. The most remarkable difference between the two graphite grades is the concentration of QI particles. Although the QI particles are densified after irradiation, the bonding between densified QI particles and adjacent graphite structures is week and the lenticular cracks within the adjacent graphite structures were not effectively closed (Fig. 8). The QI particles should shrink on their own during irradiation. Therefore, it is believed that the densification of the QI particles does not contribute obviously to the shrinkage of graphite during irradiation.

The volume changes of NBG-18 and IG-110 during neutron irradiation at 750 °C has been revealed recently [1]. Given the scattering of the irradiation data, the maximum dimensional shrinkage of the NBG-18 graphite and IG-110 graphite can be estimated to be -4.5% and -5% respectively. The NBG-18 graphite shows a smaller maximum volume shrinkage than IG-110 graphite, which is suggested to be a result of a much higher concentration of QI particles in NBG-18 graphite.

## 5. Conclusions

One typical microstructure of nuclear graphite, the QI particle, was investigated in this study. The QI particles were found to evolve into hillocks on the irradiated surface of graphite, making ion irradiation combined with SEM characterization an effective technique for studying QI particle's distribution and concentration in various nuclear graphites. The shrinkage of a QI particle was directly observed during in-situ electron irradiation in a TEM. The TEM characterization of xenon-ion-irradiated NBG-18 graphite shows that the QI particles evolved into densified spheres which have weak bonding with the surrounding graphite structures. It is believed that the densification of QI particles does not contribute obviously to graphite volume shrinkage during irradiation. High-temperature xenon ion irradiation induced a continues increase of the Raman G peak width of graphite. However, at doses > 15 dpa, both G peak width and $I_D/I_G$ ratio tended to be stable, indicating a stable graphitic structure (although damaged) in nuclear graphite. The concentration of QI particles in NBG-18 is characterized to be much higher than that in IG-110, which is suggested to be the reason why NBG-18 has a smaller maximum dimensional shrinkage than IG-110 during irradiation.

## Acknowledgments

This work was supported by Youth Innovation Promotion Association of the Chinese Academy of Sciences (grant number: 2019262); and the National Natural Science Foundation of China (grant number: 11505265, 11805256, 11805261).